\documentclass[amssymb,aps,showpacs,twocolumn]{revtex4}
\usepackage[]{graphicx}
\usepackage[]{amsmath}
\usepackage{hyperref}

\def\ket#1{| #1 \rangle}

\def\II{1\!\mathrm{l}}

\def\cC{\mathcal{C}}
\def\cE{\mathcal{E}}
\def\cG{\mathcal{G}}

\def\cL{\mathcal{L}}

\def\cN{\mathcal{N}}
\def\cP{\mathcal{P}}

\def\cS{\mathcal{S}}
\def\cT{\mathcal{T}}

\def\bL{{\bf L}}

\def\bs{{\bf s}}

\begin{document}

\title{Optimal and Efficient Decoding of Concatenated Quantum Block Codes}
\author{David Poulin}
\affiliation{Center for the Physics of Information, California Institute of Technology, Pasadena, CA 91125}

\date{\today}

\begin{abstract}
We consider the problem of optimally decoding a quantum error correction code --- that is to find the optimal recovery procedure given the outcomes of partial ``check" measurements on the system. In general, this problem is NP-hard. However, we demonstrate that for concatenated block codes, the optimal decoding can be efficiently computed using a message passing algorithm. We compare the performance of the message passing algorithm to that of the widespread blockwise hard decoding technique. Our Monte Carlo results using the 5 qubit and Steane's code on a depolarizing channel demonstrate significant advantages of the message passing algorithms in two respects.
1)~Optimal decoding increases by as much as 94$\%$ the error threshold below which the error correction procedure can be used to reliably send information over a noisy channel. 2)~For noise levels below these thresholds, the probability of error after optimal decoding is suppressed at a significantly higher rate, leading to a substantial reduction of the error correction overhead.
\end{abstract}


\maketitle

\section{Introduction}

Quantum error correction (QEC)~\cite{QEC} and fault-tolerant quantum computation~\cite{FT} demonstrate that quantum information can in principle be stored and manipulated coherently for arbitrarily long times despite the presence noise. The general framework of QEC is the following. Redundancy is introduced by encoding the information of system $S$ into a larger system $S'$. The image of $S$ in $S'$ characterizes a code, while a particular embedding of $S$ into $S'$ is called an encoding. The system $S'$ is subjected to some noise. Partial measurements whose outcomes are known as the``error syndrome" are performed on $S'$. Conditioned on this error syndrome, a recovery operation is applied to $S'$ in order to restore its original information. This last step, called ``decoding", is the subject of the present study.

In the absence of structure in the code, we know from a classical result~\cite{BMT78a} that finding the optimal recovery is NP-hard. For practical purposes, one must either use codes with lots of structure --- which typically offer poorer performances --- or settle for suboptimal recovery. Residual errors after
decoding are therefore of two varieties: those due to the information-theoretic limitations of the code and those arising from suboptimal decoding procedures. In the past decades, considerable progress has been made towards understanding this tradeoff in the classical setting (see e.g. \cite{Mac03a,RU05a} and references therein). Central to these advancements is the use of the message passing decoding algorithm pioneered by Gallager \cite{Gal63a} which often leads to near-optimal decoding. This technique was recently introduced in the quantum realm by Ollivier et al. \cite{OT05b, COT05a} for the decoding of low density parity check (LDPC) codes (see also \cite{MMM04a,OT05a} for related work).

Concatenation of block codes is widely used in quantum information science and is a key component of almost all fault tolerant schemes (a noticeable exception is topological quantum computing~\cite{Kit03a}). As the name suggests, the system $S'$ that redundantly encodes the information of system $S$ can itself be encoded in a yet larger system $S''$, adding an extra layer of redundancy. Provided the initial error rate is below a threshold value, every extra level of concatenation should reduce the probability of error after decoding, so concatenation can in principle be repeated until the error is below any desired value. 

In this article, we demonstrate an efficient \footnote{Our notion of efficient decoding differs to that of Ref. \cite{FSW06a}: our decoding runs in linear time with the number of physical qubits.} message passing algorithm that achieves optimal (maximum likelihood) decoding for {\em concatenated block codes} with uncorrelated noise. We numerically investigate the message passing algorithm using the 5 qubit code \cite{5qubit} and Steane's 7 qubit code \cite{Ste96a} and compare their performances to the commonly used blockwise minimal-distance decoder (based on a local rather than global optimization). The advantages of the message passing algorithm are substantial. On the one hand, for the 5 qubit code used on a depolarizing channel, the message passing algorithm can correctly decode the information for a noise level up to at least 0.1885 (the exact threshold is probably the hashing bound $\approx 0.189$) compared to the values 0.1376 previously established using blockwise decoding \cite{RDM02a}. For Steane's code, this enhancement is even greater going from 0.0969 \cite{RDM02a} to at least 0.188. On the other hand, away from these noise thresholds, the probability of error decreases at a significantly higher rate using optimal decoding. For instance, for a 0.1 depolarizing channel and using 4 levels of concatenation of the 5 qubit code, the probabilities that the blockwise decoding and the optimal decoding fail to correctly identify the error differ by more that 3 orders of magnitude. As a consequence, a decoding error probability $p_e \leq \delta$ for any $\delta>0$ can be achieved with a substantially reduced error correction overhead. 

\section{Stabilizer formalism} 

Our presentation of the stabilizer formalism follows \cite{Pou05b}, see \cite{Got97a} for the general theory. Denote by $X$, $Y$ and $Z$ the three Pauli matrices and by $\II$ the $2\times 2$ identity matrix. The group $\cP_1$ is the multiplicative group generated by the Pauli matrices and the imaginary unit $i$. The $n$-qubit Pauli group $\cP_n$ is the $n$-fold tensor product of $\cP_1$. We denote $X_j$ the Pauli matrix $X$ acting on the $j$th qubit for $j = 1,\ldots,n$ and similarly for $Y$ and $Z$. Note that the $X_j$'s and the $Z_j$'s are a generating set of $\cP_n$, i.e. $\cP_n = \langle i,X_j, Z_j \rangle$. The Clifford group on $n$ qubits $\cC_n$ is the largest subgroup of the unitary group $U(2^n)$ that maps $\cP_n$ to itself under the adjoint action. 

The encoding of $k$ qubits into $n$ qubits can be specified by a matrix $C \in \cC_n$. $C$ is a unitary matrix acting on $n$ qubits that are distributed in 3 different sets. The first $k$ ``logical" qubits  contain the information to be encoded in the $n$ qubits; the next $u$ ``stabilizer" qubits are set to the state $\ket 0^{\otimes u}$; and finally the remaining $r=n-k-u$ ``gauge" qubits are in arbitrary states. The image of the Pauli operators acting on the first $k$ qubits are known as logical Pauli operators $\overline X_j = CX_jC^\dagger$ and $\overline Z_j = CZ_jC^\dagger$. The image of the $Z$ Pauli operators acting on qubits $j = k+1,\ldots,k+u$ are called stabilizer generators $S_j = CZ_{k+j}C^\dagger$ whereas the image of the $X$ operators acting on those qubits are called pure errors $T_j = CX_{k+j}C^\dagger$. Finally, the image of the Pauli operators acting on the remaining $r$ qubits are called gauge operators $g_j^x = CX_{k+u+j}C^\dagger$ and $g_j^z = CZ_{k+u+j}C^\dagger$.

The stabilizer generators $S_j$ mutually commute, so can be simultaneously measured. The outcome of that measurement is called the error syndrome $s \in \{-1,1\}^u$. Since the $u$ stabilizer qubits are all in state $\ket 0$ prior to encoding, we conclude that in the absence of noise the encoded state should be a $+1$ eigenstate of all stabilizer generators, thus the error syndrome should be all ones. A non-trivial syndrome therefore indicates that an error has corrupted the register, and the task of decoding consists in finding the optimal recovery procedure given an error syndrome. 

\section{Decoding} 

To address the decoding problem, note that $\cP_n = \langle i, \overline X_j, \overline Z_j, S_j, T_j, g^x_j,g^z_j\rangle$. In other words, any element $E \in \cP_n$ can be written, up to an irrelevant phase, as 
\begin{equation}
E = \cL(E)\cT(E)\cG(E),
\end{equation}
where $\cL(E)$ is a product of logical Pauli operators, $\cT(E)$ is a product of pure errors, and $\cG(E)$ is a product of gauge operators and stabilizer elements. Moreover, this decomposition can be found by running the circuit $C$ backward, which is efficient since $C \in \cC_n$ \cite{Got97a}.  $\cT(E)$ is completely determined by the syndrome: $T_j$ appears in $\cT(E)$ if and only if the $j$th syndrome bit is $-1$. The value of $\cG(E)$ is irrelevant because the information encoded in the $n$ qubits is invariant under the action of any $\cG(E)$. This reflects the fact that the stabilizer qubits are initially set to $\ket 0$ and that the gauge qubits are in random states. Thus, to undo the effect of an error $E$, one needs to identify the most likely value of $L=\cL(E)$ given $s$, or equivalently given $T=\cT(E)$. 

For simplicity, we will focus on Pauli channels where errors $E$ are elements of $\cP_n$ distributed according to $P(E)$. Given this probability $P(E)$ over $\cP_n$ one can compute the conditional probability $P(L|T) = P(L,T)/P(T)$ using
\begin{eqnarray}
P(L,T) &=& \sum_E \delta[\cT(E) = T]\delta[\cL(E) = L] P(E) \\
&=& \sum_G P(E = LTG),
\label{eq:prob_block}
\end{eqnarray} 
where $\delta$ denotes the indicator function and $G$ takes all possible combinations of stabilizer generators and gauge operators. Given a finite block size $n$, these probabilities can be computed and the optimal decoding  $\hat L(T) = \mathrm{argmax}_L\{P(L|T)\}$ can be evaluated. Decoding a block code thus consists of looking in a table containing the values of $\hat L(T)$ for each $T$. Typically --- and in particular for a non-degenerate code over the depolarization channel --- $\hat L(T)$ corresponds to the minimal distance decoder $\cL(\hat E(T))$ where $\hat E(T)$ is the error acting on the fewest number of qubits and that is compatible with the observed syndrome. 

Concatenation is realized by encoding the $n$ qubits of the code in an other code. There is no need for this other code to be identical to the original one. However to simplify the presentation, we will  assume that the same code is used at every concatenation layer and that it encodes a single qubit in $n$ qubits; generalizations are straightforward. This procedure can be repeated $\ell$ times at the expense of an exponentially growing number of physical qubits $n^\ell$. The number of stabilizer generators grows roughly as $un^{\ell-1}$ (it is a geometric sum); thus the syndrome can take $2^{un^{\ell-1}}$ different values. Thus, even for moderate values of $\ell$, it is not feasible to construct a lookup table giving the optimal decoding procedure for each syndrome value.

What is generally done to circumvent this double exponential blowup is to apply the optimal recovery {\em independently for each concatenation layer} (see e.g. \cite[Chap. 6]{Got97a} and references therein). One first measures the syndrome from each of the $n^{\ell-1}$ blocks of $n$ qubits of the last layer of concatenation, and optimally decodes them using the lookup table. One then moves one layer up and applies the same procedure to the $n^{\ell-2}$ blocks of the second-to-last layer, etc. When the initial error rate is below a certain threshold value, the probability $p_e$ that this procedure fails to correctly identify $\cL(E)$ decreases doubly-exponentially with $\ell$. Hence, this decoding scheme based on hard decisions for each concatenation layer is efficient, leads to a good error suppression, but is nonetheless suboptimal. 

\subsection{Optimal decoding}

\begin{center}
\includegraphics[width=2.9in]{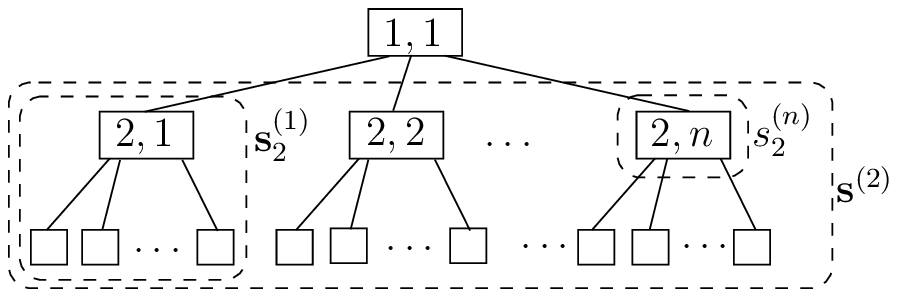}
\end{center}
Let $s_m^{(j)} \in \{-1,1\}^u$ be the syndrome of the $j$th block of the $m$th concatenation layer. Denote $\bs^{(j)}_m$ the collection of syndromes whose stabilizers act non-trivially on the physical qubits associated to the $j$th block of the $m$th concatenation layer: these sets can be defined recursively by $\bs^{(j)}_m = \{s_m^{(j)}\} \cup \{\cup_{i=jn-j+1}^{jn} \bs_{m+1}^{(i)}\}$ with the initialization $\bs^{(j)}_\ell = s^{(j)}_\ell$. Finally, denote $\bs_m = \cup_{j=1}^{n^{m-1}} \bs_m^{(j)}$ all the syndromes from the layers $m$ to $\ell$. (See the above figure for a pictorial representation of $s_m^{(j)}$, ${\bf s}_m^{(j)}$, and ${\bf s}_m$.) Then, $\bs_1$ is the set of all syndromes and maximum likelihood decoding consists in finding $\mathrm{argmax}_{L_1} P(L_1|\bs_1)$. This probability can be factorized by conditioning on the logical errors of the second layer $\bL_2 = (L_2^{(1)},\ldots,L_2^{(n)})$:
\begin{widetext}
\begin{eqnarray}
P(L_1|\bs_1) 
&=& \sum_{\bL_2} P(L_1|\bs_1,\bL_2) P(\bL_2|\bs_1) \label{eq:factor} \\
&=& \sum_{\bL_2} \delta[L_1 =  \cL(\bL_2)] 
\frac{P(\bL_2,\bs_1)}{P(\bs_1)} \nonumber \\
&=& \sum_{\bL_2} \delta[L_1 =  \cL(\bL_2)] 
\frac{P(s_1|\bL_2,\bs_2)P(\bL_2,\bs_2)}{P(s_1,\bs_2)} \nonumber \\
&=& \sum_{\bL_2} \delta[L_1 =  \cL(\bL_2)] \delta[s_1 = \cS(\bL_2)]
\frac{P(\bL_2|\bs_2) P(\bs_2)}{P(s_1,\bs_2)} \nonumber\\
&=& \sum_{\bL_2} \frac{\delta[L_1 =  \cL(\bL_2)] \delta[s_1 = \cS(\bL_2)]}{P(s_1|\bs_2)}
\prod_{j=1}^n P(L^{(j)}_2|\bs^{(j)}_2). \nonumber
\end{eqnarray}
\end{widetext}
Above, $\cS(\bL)$ denotes the syndrome associated to the error pattern $\bL \in \cP_n$. This series of manipulations repeatedly uses Bayes' rule and the fact that the syndrome and logical error of level $m$ are completely determined given the logical errors of layer $m+1$.  The last step relies on the important assumption that the channel is memoryless, or more specifically, that the noise model does not correlate qubits across distinct blocks (errors on qubits in the same block could be correlated).  

\begin{center}
\includegraphics[width=2.6in]{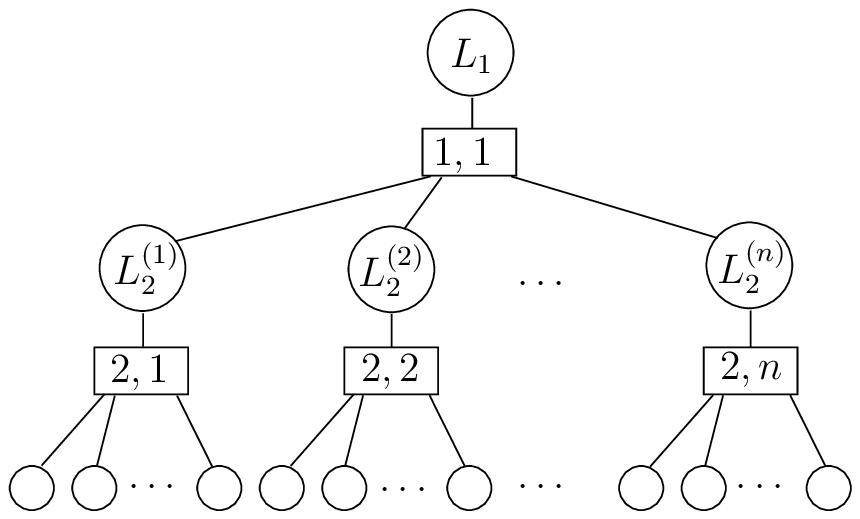}
\end{center}

Equation~(\ref{eq:factor}) shows that by conditioning on the logical errors of each concatenation layer, the {\em factor graph} associated to the function $P(L_1|\bs_1)$ is a tree, as depicted in the above graph. 
We have thus reduced optimal decoding to a {\sc sum product} problem (known as tensor network contraction in quantum information science \cite{MS05a}) on a tree graph which can be solved exactly and efficiently in the number of variables using a message passing algorithm (also known as belief propagation); see \cite{AM00a, Mac03a, RU05a}, and references therein. Let us describe this algorithm in a general setting. 

The factor graph is a bipartite graph, and vertices from the two partitions are decorated with circles and boxes. Circle vertices are labeled $c=1,\ldots,N$ and each one carries a variable $x_c$ with value in a discrete set. Box vertices are labeled $b = 1,\ldots,M$, and each one contains a function $f_b$ that depends on the variables $x_c$ from the adjacent circles $c \in \cN(b)$, collectively denoted $X_b = \{x_c:c \in \cN(b)\}$. The goal is to compute marginals 
\begin{equation}
f(x_c) = \frac 1Z \sum_{\{x_1,\ldots x_N\}\backslash x_c} \prod_{b=1}^M f_b(X_b),
\end{equation}
where $\backslash x_c$ indicates that $x_c$ is omitted from the set and $Z$ is a normalization factor. To this end, messages $q_{c\rightarrow b}$ are passed from the circles to the boxes and messages $r_{b \rightarrow c}$ are passed from the boxes to the circles following the rules
\begin{eqnarray}
q_{c\rightarrow b}(x_c) &=& \prod_{b' \in \cN(c)\backslash b} r_{b' \rightarrow c}(x_c) \\
r_{b\rightarrow c}(x_c) &=& \sum_{X_b\backslash x_c} \left[f_b(X_b) \prod_{c' \in \cN(b)\backslash c} q_{c'\rightarrow b}(x_{c'})\right],
\end{eqnarray}
where $\cN(b)\backslash c$ means all neighbors of $b$ excluding $c$, and similarly for $\cN(c)\backslash b$. Note that these messages are functions of the discrete variables $x_c$ (i.e. they are vectors). The $q_{c\rightarrow b}$ messages are initialized to the constant function 1. For a tree graph, the desired marginal is obtained from these messages after a number of steps equal to the depth of the variable $x_c$ and is given by $f(x_c) = k\prod_{b\in \cN(c)} r_{b\rightarrow c}$, where $k$ is a normalization factor.

In the case of interest, circles carry logical operators and a box labeled $m,j$ carries the function $\delta[L_{m+1}^{(j)} = \cL(L_m^{(nj-n+1)},\ldots,L_m^{(nj)})]\delta[s_{m+1}^{(j)} = \cS(L_m^{(nj-n+1)},\ldots,L_m^{(nj)})]$, where the syndrome is fixed by the measurements. To complete the picture, extra box vertices carrying the function Eq.~(\ref{eq:prob_block}) need to be attached to the bottom leaves of the graph. The factor $p(s_1|\bs_2)^{-1}$ can be evaluated by normalizing the obtained distribution. Thus, we can efficiently evaluate $P(L_1|{\bf s}_1)$ \footnote{As presented here, the algorithm allows to simultaneously compute $P(L_m^{(j)})$ for all $m,j$. Computing only $P(L_1|{\bf s}_1)$ leads to simplification of the message passing rules; in particular, messages need only to flow from the bottom to the top of the graph.}, and the optimal recovery is the $L_1$ maximizing this function. 

The advantage of the message passing algorithm over the minimal distance decoder comes from the fact that it does not throw away useful information \cite{For66b}. Instead of computing the most likely recovery and passing it on to the next level of coding, the entire list of probability of possible recoveries, conditioned on the observed syndrome, is passed on. In other words, the original channel is composed with the syndrome measurement, and projected onto the logical algebra to yield a ``conditionally renormalized" channel. 

\section{Numerical results} 

Following the tradition for benchmarking QEC techniques, we investigate the performance of the message passing decoding algorithm using a depolarization channel, where each qubit is independently subjected to the channel
\begin{equation}
\cE_p(\rho) = (1-p)\rho + \frac p3 \left( X\rho X+Y\rho Y + Z\rho Z \right).
\end{equation}
We use the 5 qubit code \cite{5qubit} concatenated with itself up to $\ell = 10$ times, for an overhead of 9,765,625  physical qubits per logical qubit. Pauli errors $E \in \cP_{n^\ell}$ are generated by picking each $n$ single-qubit operator independently according to the probability $P(\II) = 1-p$, $P(X) = P(Y) = P(Z) = p/3$. The associated logical error $\cL(E)$ and syndromes $\cS(E)$ are computed exactly. These syndromes are used by a blockwise decoding routine yielding an estimate $L_{BW}$ and by a message passing routine yielding the optimal decoding $\hat L$. A decoding is declared incorrect when its estimate differs from $\cL(E)$. This is repeated a large number of times ($10^4-10^8$) to evaluate the probability $p_e$ that the decoding gives an incorrect estimate. 

\begin{figure}[tbh] 
\center \includegraphics[height=2.1in]{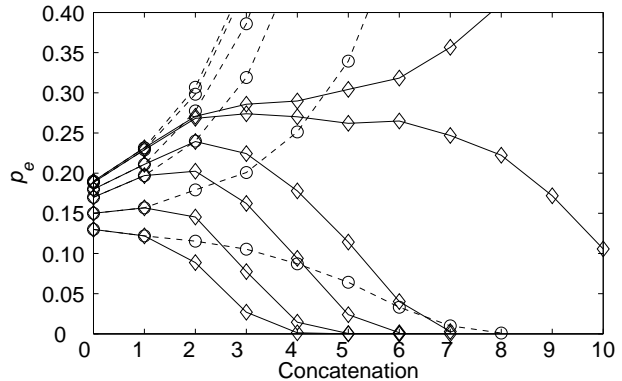}
\caption{Monte Carlo results for the 5 qubit code showing the probability of erroneous decoding $p_e$ as a function of the level of concatenation $\ell$ for different depolarization rate $p = 0.13,\ 0.15,\ 0.17,\ 0.18,\ 0.1885,$ and $0.19$. The diamonds are from the message passing algorithm, and the circles are from the blockwise decoding. All data are from samples of $2\cdot 10^4$ encoded qubits.}
\label{fig:flow}
\end{figure} 
Figure~\ref{fig:flow} shows the probability of incorrect decoding $p_e$ for both the blockwise and the optimal decoding as a function of the level of concatenation $\ell$ and for different channel parameters $p$ ranging from 0.13 to 0.19. For the blockwise decoding, $p_e$ ceases to decrease with $\ell$ for values of $p \geq 0.15$. This reflects the fact  that the threshold of this decoding technique for this particular code is about 0.1376 \cite{RDM02a}, so all curves except the 0.13 one are above the threshold. On the other hand, optimal decoding succeeds in decreasing the error probability for values of $p$ up to at least 0.1885, but appears to fail at $p=0.19$. We conjecture that the exact value of this threshold is the hashing bound $\simeq 0.189$, where the single-qubit coherent information vanishes and is the highest threshold any non degenerate code can achieve \cite{DSS98a}. Results obtained from Steane's code \cite{Ste96a} show a quite similar behavior, with at least 94$\%$ increase of threshold going from 0.0969 \cite{DSS98a} to at least 0.188 and appears to fail at 0.1885.     

An interesting feature of the $p_e(\ell)$ curves obtained from optimal decoding is their non-monotonicity. Blockwise decoding, on the other hand, always yields monotonic curves for this type of channel; thus its global behavior under concatenation can be predicted from a single level of coding. This is because decoding is performed independently on each concatenation layer. With the optimal decoder, information about the syndromes is propagated from one layer of concatenation to the next through the conditionally-renormalized channel that ceases to be depolarizing and varies from one qubit to the other. Thus, non-monotonicity of the $p_e(\ell)$ curves is a signature of the global optimization performed by the message passing algorithm. 

\begin{figure}[t!]
\center \includegraphics[height=2.1in]{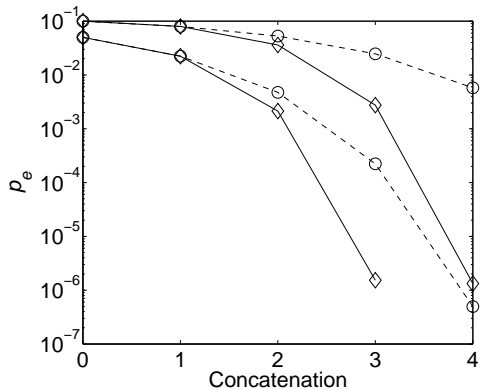}
\caption{As in Fig.\ref{fig:flow} for $p=0.1$ and 0.05. Diamonds obtained from samples of $10^8$ encoded qubits. Circle were produced using an exact numerical technique similar to that of Ref.~\cite{RDM02a}.}
\label{fig:exponential}
\end{figure} 
Figure~\ref{fig:exponential} shows the behavior of $p_e$ as a function of $\ell$ away from the threshold values, i.e. in the natural operating regime of the code. Again, the advantages of the message passing algorithm are considerable. After 4 rounds of concatenation for $p=0.1$, message passing fails with a probability of roughly $10^{-6}$, whereas this probability is well above $10^{-3}$ for blockwise hard decoding. It takes $6$ layers of concatenation for the blockwise decoding to reach comparable performances. Again, results obtained from Steane's code show an even larger gap.

Finally, we once again stress that the message passing outputs the probability of an error $L$ rather than a particular value of $L$. A hard decision can then be made based on this probability. We observe that when decoding succeeds, $P(\hat L)$ is typically very close to one (e.g. 0.999 for $\ell = 3$) whereas when it fails it is relatively low (typically 0.7); the algorithm knows that it is failing. This ``flagging" of errors offers a great advantage when post-selection is an option. The possibility of operating the algorithm with soft inputs, i.e. noisy syndrome measurements, is also of interest in several circumstances. 

\section{Conclusion} 

We have demonstrated an efficient message passing algorithm for the optimal decoding of concatenated quantum block codes on a memoryless channels. Numerical results show substantial benefits of our approach over the widely used blockwise hard decoding, including an increase of error thresholds and a greater error suppression rate. Message passing algorithms have been used on graphs with loops (describing e.g. LDPC codes, turbo-codes, or channels with memory) and often yield near-optimal decoding. The quantum generalization of these schemes, including quantum LDPC codes \cite{MMM04a,COT05a} and quantum turbo-codes \cite{OPT06a}, are promising avenues for the realization of a quantum information technologies. Techniques reminiscent of message passing have been used to beat the hashing bound but were not efficiently implementable~\cite{DSS98a,SS06a}: efficient decoding may now be within reach using our techniques. A ``hard" message passing scheme was also used in \cite{Kni05a} to obtain high fault-tolerant error thresholds: a full-fledge message passing scheme --- although not optimal for correlated errors that are typically present in fault-tolerant schemes --- should further improve this threshold and may significantly reduce the resource overhead.

I thank Harold Ollivier for several useful conversations on message passing algorithms and quantum error correction, and Graeme Smith and Jon Yard for comments on this paper. This work was supported in part by the Gordon and Betty Moore Foundation through Caltech's Center for the Physics of Information, by the National Science Foundation under Grant No. PHY-0456720, and by Canada's NSERC.



\end{document}